# Structural, magnetic, electric, dielectric, and thermodynamic properties of multiferroic GeV$_4$S$_8$

S. Widmann,[1] A. Günther,[1] E. Ruff,[1] V. Tsurkan,[1,2] H.-A. Krug von Nidda,[1] P. Lunkenheimer,[1] and A. Loidl[1,*]

[1]*Experimental Physics V, Center for Electronic Correlations and Magnetism, Institute of Physics, University of Augsburg, 86135 Augsburg, Germany*
[2]*Institute of Applied Physics, Academy of Sciences of Moldova, Chisinau MD-2028, Republic of Moldova*

*alois.loidl@physik.uni-augsburg.de

**Abstract**

The lacunar spinel GeV$_4$S$_8$ undergoes orbital and ferroelectric ordering at the Jahn-Teller transition around 30 K and exhibits antiferromagnetic order below about 14 K. In addition to this orbitally driven ferroelectricity, lacunar spinels are an interesting material class, as the vanadium ions form V$_4$ clusters representing stable molecular entities with a common electron distribution and a well-defined level scheme of molecular states resulting in a unique spin state per V$_4$ molecule. Here we report detailed x-ray, magnetic susceptibility, electrical resistivity, heat capacity, thermal expansion, and dielectric results to characterize the structural, electric, dielectric, magnetic, and thermodynamic properties of this interesting material, which also exhibits strong electronic correlations. From the magnetic susceptibility, we determine a negative Curie-Weiss temperature, indicative for antiferromagnetic exchange and a paramagnetic moment close to a spin $S = 1$ of the V$_4$ molecular clusters. The low-temperature heat capacity provides experimental evidence for gapped magnon excitations. From the entropy release, we conclude about strong correlations between magnetic order and lattice distortions. In addition, the observed anomalies at the phase transitions also indicate strong coupling between structural and electronic degrees of freedom. Utilizing dielectric spectroscopy, we find the onset of significant dispersion effects at the polar Jahn-Teller transition. The dispersion becomes fully suppressed again with the onset of spin order. In addition, the temperature dependencies of dielectric constant and specific heat possibly indicate a sequential appearance of orbital and polar order.

## I. INTRODUCTION

GeV$_4$S$_8$ belongs to the large material class of lacunar spinels, ternary chalcogenides of composition $AM_4X_8$ ($A$ = Ga and Ge; $M$ = V, Mo, Nb, and Ta; $X$ = S and Se), which exhibit the fcc GaMo$_4$S$_8$ structure at room temperature and were first synthesized by Barz [1] half a century ago. Later on, these compounds were studied in detail by Perrin *et al*. [2,3], Brasen *et al*. [4], and Vandenberg and Brasen [5]. Lacunar spinels can be derived from the normal spinel structure $AM_2X_4$ by removing every second $A$-site ion from its tetrahedral site. In addition, structural rearrangements including slight shifts of anions and cations lead to the formation of regular molecular units. Unusual electronic ground-state properties were explained by the fact that weakly linked molecular units, namely cubane ($M_4X_4$)$^{n+}$ and tetrahedral ($AX_4$)$^{n-}$ clusters are the building blocks of lacunar spinels. These molecular units are arranged in a fcc type lattice and can be described by a unique electron density and a well-defined molecular orbital scheme [6]. Astonishingly, in contrast to the ferromagnetism found in the majority of these compounds [1,4], semiconducting GeV$_4$S$_8$ exhibits an antiferromagnetic ground state [7,8].

In most cases, the cubane-like metallic clusters are characterized by an only partly filled electronic shell and, hence, most of the lacunar spinels are Jahn-Teller active [8]. Recently, for GeV$_4$S$_8$ [9] and GaV$_4$S$_8$ [10] it has been shown that the onset of orbital order and the resulting Jahn-Teller distortion induces ferroelectricity, and, thus, these compounds are multiferroic. The polar dynamics in GaV$_4$S$_8$ at the phase transition has been probed by dielectric and THz spectroscopy [11]. In addition, lacunar spinels are dominated by Mott physics and a plethora of correlation effects has been reported so far: d-derived heavy-fermion behavior [12,13], pressure-induced superconductivity [14], bandwidth-controlled metal-to-insulator transitions [15,16], large negative magnetoresistance [17], a two-dimensional topological insulating state [18], and resistive switching via an electric-field induced transition [19,20,21]. Further excitement concerning this material class has been created by the fact that GaV$_4$S$_8$ exhibits a complex magnetic phase diagram, with cycloidal and Néel-type skyrmion phases, which are embedded in the ferromagnetic ground state [22,23]. In GaV$_4$S$_8$, the skyrmion lattice and even single magnetic skyrmions were found to be dressed with ferroelectric polarization [10].

Here we report the synthesis and detailed characterization of large single crystals of GeV$_4$S$_8$. The V$_4$ clusters in GeV$_4$S$_8$ are occupied by 8 electrons, leaving two unpaired electrons with total spin $S = 1$ in the highest excited triplet [7,8]. As a consequence, one hole in the triplet state makes this compound Jahn-Teller active and GeV$_4$S$_8$ undergoes a structural phase transition from the room temperature $F\bar{4}3m$ structure into an orthorhombic and orbitally ordered phase (*Imm*2) at $T_{JT} \approx$ 30 - 33 K [9,24,25]. In the orbitally ordered phase, the V$_4$ molecular tetrahedra distort with one short and one long V-V bond on adjacent sites [25], thereby inducing long-range ferroelectric order [9]. Subsequently, antiferromagnetic (AFM) order is established at $T_N \approx$ 13 - 18 K [7,8,9,24,25], and the symmetry of the magnetically ordered phase was described by the space group $P_b mn2_1$ [8]. It has been reported that the characteristic ferroelectric distortion of the vanadium cubane units appearing at the Jahn-Teller transition is even enhanced when entering the antiferromagnetic ground state [8]. Later on, it was concluded that the onset of antiferromagnetic ordering happens isostructurally without any symmetry changes but only by an increase of the existing distortion [25]. It also seems that the charge distribution, which by symmetry is regular among the four vanadium ions at room temperature, changes significantly, leading to different localized magnetic moments at the adjacent vanadium sites with short, respectively long bonds [8]. It has been speculated that the difference in localized moments results from charge-ordering phenomena reminiscent to the Verwey transition in magnetite [26]. We thought it would be worthwhile to unravel a number of open questions and to further investigate the interesting physics of GeV$_4$S$_8$. Consequently, in the course of this work we grew single crystals and characterized them by detailed electric, magnetic, thermodynamic, thermal expansion, and dielectric experiments, to study in detail the orbitally ordered state and the nature of the ferroelectric and antiferromagnetic phase transitions in this compound.

## II. EXPERIMENTAL DETAILS

Polycrystalline GeV$_4$S$_8$ was prepared by solid-state reaction using pure elements of Ge (6N), V (3N), and S (5N). Three subsequent synthesis steps were necessary to obtain complete reaction of the starting materials and to form the stoichiometric ternary phase. After each step, phase purity was checked by x-ray powder diffraction and polycrystals were used as starting material for single-crystal growth using chemical transport reactions. The crystals were grown in closed quartz ampoules at temperatures between 800 and 900 °C utilizing iodine as transport agent. Typical crystals grown in the course of this work are shown in Fig. 1. The dark shiny crystals with metallic luster have the shape of truncated octahedrons and plates with size up 5×3×2 mm$^3$. The largest mirror-like planes are <111> planes of the room-temperature fcc lattice.

Magnetic measurements were performed with a SQUID magnetometer (Quantum Design MPMS XL) in the temperature range from 1.8 K < $T$ < 400 K and in external magnetic fields up to 5 T. The electrical resistivity was measured via a standard four-point measuring technique using the dc option of a Physical Properties Measurement System (Quantum Design PPMS). At low temperatures, where the samples exhibit large resistivity, additional measurements in two-point configuration with a Keithley 617 electrometer and excitation voltages up to 100 V have been performed. The crystals for the resistivity measurements were thin platelets with an area of approximately 3 mm² and a thickness of 1 mm. All contacts were made by silver paint with current and voltage in most cases along the crystallographic <111> direction.

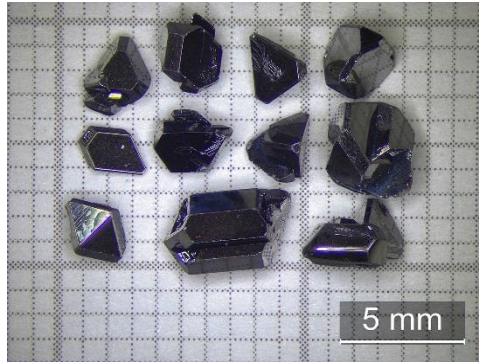

FIG. 1. GeV$_4$S$_8$ single crystals grown and characterized in the course of this work, placed on a 1 mm grid.

Standard heat capacity was investigated in a Quantum Design PPMS for temperatures 1.8 K < $T$ < 300 K and in magnetic fields up to 9 T. To identify the nature of the structural and magnetic phase transitions, the specific-heat experiments were supplemented by a large-pulse method, applying heat pulses leading to a ca. 20% increase of sample temperature. The temperature response of the sample was thereby separately analyzed on heating and cooling. In a second-order phase transition, both responses should be equal, while significant differences are expected at first-order phase transitions. The dielectric properties of GeV$_4$S$_8$ were measured at audio and radio frequencies for temperatures between 4 and 300 K. These experiments in the frequency range from 1 Hz to 10 MHz were performed with a frequency-response analyzer (Novocontrol Alpha-Analyzer). In the dielectric experiments, we used platelet-shaped single crystals of approximately 4 mm² cross section and 0.4 mm thickness, in most cases with the electric field applied along the crystallographic <111> direction. For this purpose, silver-paint contacts were applied to opposite sides of the crystals.

## III. RESULTS AND DISCUSSION

### A. Structural details

Figure 2 shows the x-ray powder diffraction pattern of crushed single crystals, which was fitted using Rietveld refinement. The fit provides a good description of the scattered intensities. However, close to scattering angles of 35° and 43°, small and unidentified extra Bragg reflections can be detected with intensities above experimental uncertainty, which in Fig. 2 are marked by asterisks. We were not able to identify the chemical composition and possible crystallographic structure of this impurity phase. As similar extra Bragg intensities also appear in the ceramic samples, which were produced during the synthesis cycles of the sample preparation, we speculate that these extra reflections could correspond to a superstructure of vacancies. Again we were not able to identify the possible symmetry of this superstructure and no similar observations have been reported in the published literature on lacunar spinels. At room temperature we found the correct GaMo$_4$S$_8$-type crystal structure with $F\bar{4}3m$ symmetry and a lattice constant $a$ = 0.9667 nm, in good agreement with the findings of Ref. [7], reporting a room-temperature value of 0.9655(1) nm.

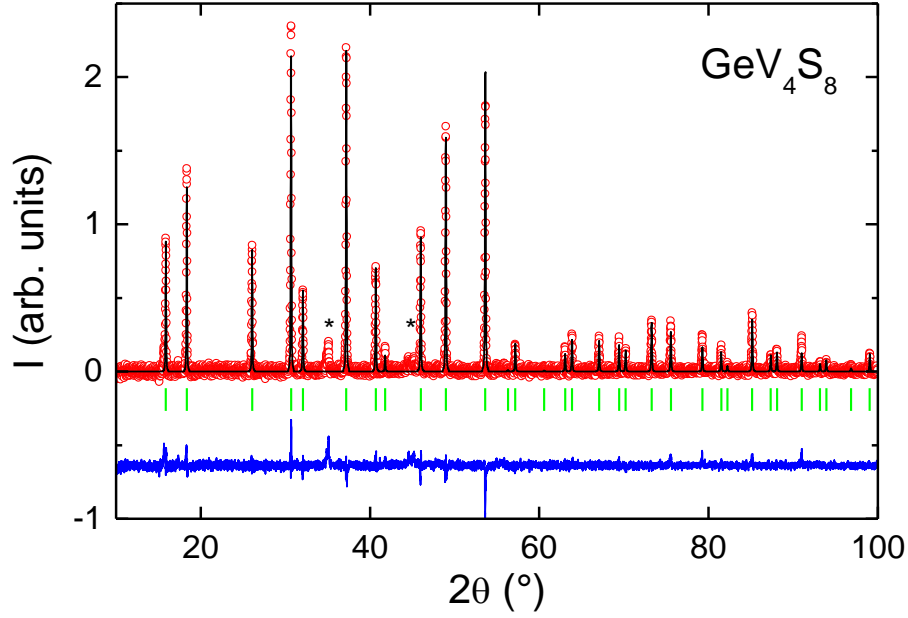

FIG. 2. X-ray diffraction pattern (open solid circles) and Rietveld refinement (solid lines) for GeV$_4$S$_8$ at room temperature. The vertical bars indicate Bragg positions of the fcc lattice. The solid line at the bottom of the figure indicates the difference pattern between observed and calculated intensities. Small extra Bragg peaks of unknown origin are detected close to 35° and 43° and are marked by asterisks.

## B. Electrical resistivity

The electrical dc resistivity in GeV$_4$S$_8$ was measured between 400 and 5 K (Fig. 3). In this temperature range the resistivity varies by almost 14 orders of magnitude. For these measurements, we have combined standard four-point dc current experiments using constant current and two-point measurements, utilizing electrometers with constant voltages from 0.1 up to 100 V. Down to the lowest temperatures, the electrical resistivity steadily increases and reaches values exceeding 100 TΩcm. The hump-like shape in $\rho(T)$, which occurs close to 100 K, most probably indicates the transition from Arrhenius-like behavior at high temperatures to variable-range-hopping (VRH) conductivity at low temperatures as discussed below. In addition, at low temperatures, we found an extreme voltage dependence. Thus is indicated in Fig. 3, where we compare measurements at 1 V (squares) with those at 10 V (circles). At 20 K, the voltage dependence of the resistivity between 1 and 10 V amounts approximately 2 decades. The strong voltage dependence in GeV$_4$S$_8$ at low temperatures probably is an intrinsic material property and does not represent the influence of depletion layers resulting from the electrical contacts. In lacunar spinels, including the system under investigation, strong electric-field induced resistive switching effects have been detected, reaching colossal values at low temperatures [27].

In the resistivity measurements presented in Fig. 3, the structural as well as the magnetic phase transitions can clearly be detected. If we analyze the slope of the electrical resistivity in Arrhenius representation for 100 < $T$ < 300 K, as shown in the right inset of Fig. 3 and if we assume purely intrinsic semiconducting behavior, $\rho_{dc} \propto \exp(E_g/2k_BT)$, we arrive at a value of the gap energy $E_g \approx 0.3$ eV. This value is of the order of the band gap determined in GaV$_4$S$_8$ [23] and as typically observed in lacunar spinels. Most of the compounds investigated so far were characterized as narrow-gap Mott insulators with electronic band gaps in the 0.2 – 0.3 eV range [27]. In LDA + U calculations for GeV$_4$S$_8$ performed by Müller *et al*. [8] the electronic band gap has been calculated to be of order 0.2 eV.

In a limited temperature range from approximately 10 to 40 K, a logarithmic plot of the resistivity vs. $T^{-1/4}$ provides a linearization of the experimental results (left inset in Fig. 3). This fact points towards a variable-range hopping (VRH) mechanism, which implies phonon-assisted tunneling of charge carriers between localized states as the dominant transport mechanism in this temperature regime. The resistivity can be described by $\rho \propto \exp(T_0/T)^{1/4}$ with a parameter $T_0 = 1.2 \times 10^7$ K. Within the VRH model, this parameter is related to the spatial extension of localized states, and, e.g., $T_0$ values of the order $10^7 - 10^8$ have typically been observed for insulating manganites [28,29,30]. VRH conductivity was also reported to occur in pure GaV$_4$S$_8$ [23] and in Ga:GeV$_4$S$_8$ solid solutions [31]. In the former compound the $T_0$ parameter was reported to be about one order of magnitude lower. However, because the electrical resistivity below the Jahn-Teller transition reveals a strong voltage dependence, details of the temperature dependence should not be over-interpreted.

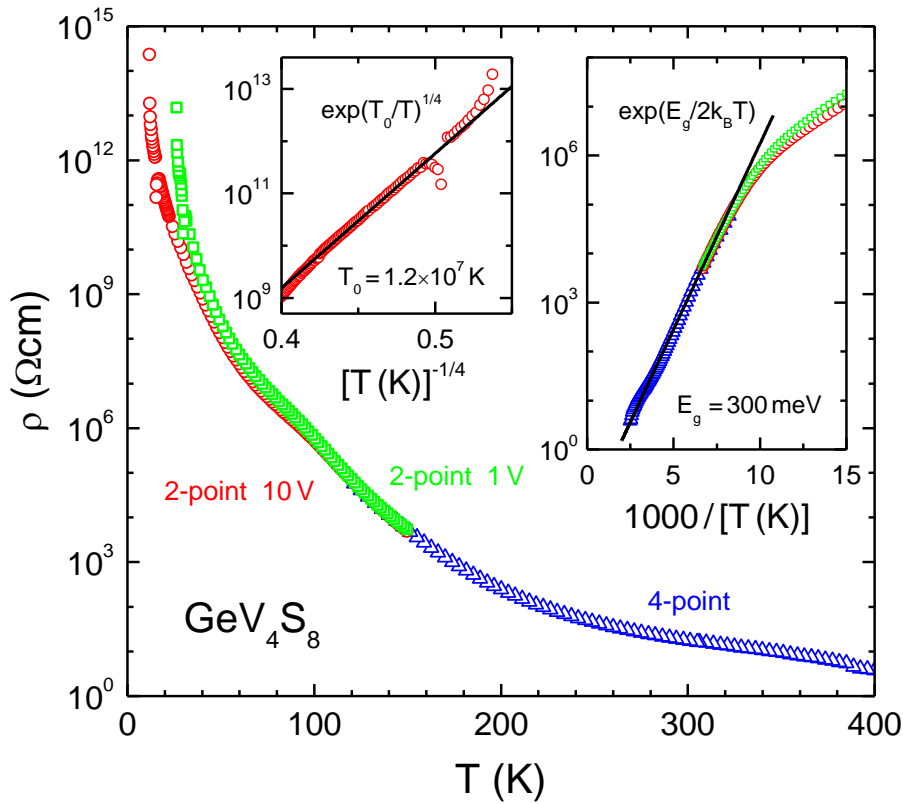

FIG. 3. Electrical dc resistivity in GeV$_4$S$_8$ for temperatures between 5 and 400 K. Triangles: Four-point constant-current results. Circles: Two-point constant-voltage results applying a voltage of 10 V. They are compared with those obtained at 1 V (squares). In the left inset, the 10 V data between about 10 and 40 K are plotted vs. $T^{-1/4}$, which should lead to a linearization of the curve for VRH behavior. The line corresponds to a linear fit. The right inset shows the resistivity data between 67 and 400 K in an Arrhenius-type representation. The solid line represents an Arrhenius fit as indicated in the inset using an energy gap of 300 meV and provides a reasonable description of the data between about 130 and 400 K.

As a concluding comment to this chapter on the electrical resistance, we would like to add that the speculation on the occurrence of charge-order, concomitant with the structural Jahn-Teller transition [8], does not seem to be supported by the dc resistivity results presented in Fig. 3. The electrical resistance is continuously increasing, revealing only a minor anomaly at the onset of orbital order. However, we have to keep in mind that the canonical Verwey transition as observed in magnetite corresponds to a metal-to-insulator transition, while in GeV$_4$S$_8$ the occurrence of charge order would correspond to a transition between two semiconducting phases.

### C. Magnetic susceptibility

Figure 4 shows the inverse magnetic susceptibility as measured in single-crystalline GeV$_4$S$_8$ in the full temperature range from 1.8 to 400 K. These experiments have been performed on cooling in external magnetic fields of 1 T with the field along <111>. At first sight, we find a well-developed antiferromagnetic (AFM) Curie-Weiss (CW) law, with two anomalies at low temperatures indicating phase transitions. From the inset of Fig. 4, it is clear that AFM order, characterized by a cusp-like anomaly, appears close to 15 K, while the structural transition, with a step-like increase of the inverse susceptibility, appears at 30 K. In literature, magnetic ordering-temperatures between 13 K and 18 K (e.g., Refs. [7,9,24,25]) and structural transitions between 30 K and 33 K [9,24,25] were reported, mostly in polycrystalline samples.

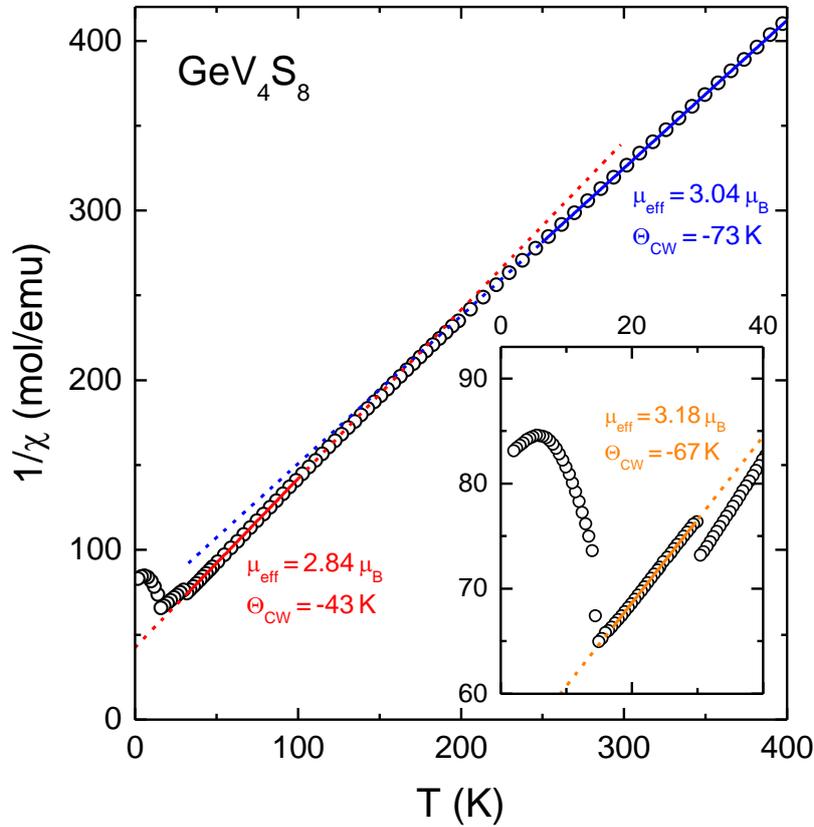

FIG. 4. Inverse susceptibility of GeV$_4$S$_8$ measured in external magnetic fields of 1 T on cooling. Solid/dashed lines indicate fits by CW laws (the solid lines indicate the fit range). The inset shows an enlarged region at low temperatures. The solid/dashed line again indicates a CW law in the temperature range between the structural and the magnetic phase transitions. The resulting fit parameters, paramagnetic moments, and CW temperatures are indicated in the figure.

An analysis in terms of a CW law for temperatures 250 < $T$ < 400 K gives a paramagnetic moment of $\mu_{eff}$ = 3.04 $\mu_B$ and a CW temperature $\theta_{CW}$ = -73 K. The inverse susceptibility shows a change of slope approximately at 150 K and an evaluation of the CW law for temperatures 30 < $T$ < 100 K results in a paramagnetic moment of 2.84 $\mu_B$ and a CW temperature of -43 K. The reason for this anomaly in the temperature dependence of magnetic exchange and paramagnetic moment is unclear at present, but could result from the onset of orbital fluctuations when approaching the Jahn-Teller transition. Below the structural phase transition at 30 K, the inverse susceptibility again follows an inverse CW law, with a paramagnetic moment of 3.18 $\mu_B$ and a CW temperature of -67 K, close to the values derived at high temperatures (see inset of Fig. 4). The values of the paramagnetic moments are close to a spin system with $S$ = 1, which gives a paramagnetic moment of 2.84 $\mu_B$, assuming a spin-only $g$-value of 2. The CW temperatures signal moderate but distinct antiferromagnetic exchange, which however is significantly enhanced when compared to the AFM phase-transition temperature $T_N$ = 15 K. This indicates competing exchange interactions or weak geometric frustration. Neither the paramagnetic moment, nor the antiferromagnetic CW temperature undergo dramatic changes at the structural phase transition. This is distinctly different from observations in ferromagnetic $GaV_4S_8$, where the CW temperature at the Jahn-Teller transition changes from antiferromagnetic to ferromagnetic [23]. CW temperatures and paramagnetic moments for $GeV_4S_8$ as determined in this work, compare roughly with reports in literature [7,24,25]. However, in these publications only one temperature range was used and the temperature ranges used for the CW fits varied considerably. It is important to notice that we found no indications of saturation effects in the inverse susceptibility up to 400 K, indicating no Van Vleck contributions from excited levels of the $V_4$ molecular cluster, as well as a stable electronic configuration of the vanadium molecules. This is in clear contrast to results in $GaV_4S_8$ where in the inverse susceptibility clear saturation effects were observed above 200 K, supporting speculations about a breakdown of stable molecular units at higher temperatures [23].

### D. Specific heat

The heat capacity of $GeV_4S_8$ has been reported by Chudo *et al*. [24] for temperatures between 1.8 and 100 K and by Bichler *et al*. [25] for temperatures up to 40 K. No detailed analysis of the heat capacity results has been performed in these investigations. However, both publications claimed a first-order nature of the structural, as well as of the magnetic transition and hysteresis effects on heating and cooling of the order of 0.5 K for the structural transition and of the order of 2.5 K for the magnetic transition have been observed [25]. We are not aware of any specific heat experiments in the compound under investigation as function of an external magnetic field. Figure 5 shows the temperature dependence of the specific heat in $GeV_4S_8$ for temperatures from 1.8 K to 300 K. For presentation purposes, we choose a semilogarithmic representation of $C/T$ with a logarithmic temperature axis. The structural and the magnetic phase transitions, both are clearly visible as lambda-like anomalies. These specific-heat experiments on single crystalline material, which were performed on cooling, exhibit anomalies at the AFM phase transition at $T_N$ = 14.6 K and at the structural transition at $T_{JT}$ = 30.5 K.

As both phase transitions seem to be of first order, we studied the structural and the magnetic phase transition also using large-pulse experiments to identify possible hysteresis effects. In analyzing cooling and heating response separately, we found significant differences in the phase transition temperatures. Depending on cooling or heating response, the heat capacity anomalies shift from 14.2 to 15.3 K at the magnetic phase transition and from 30.0 to 31.0 K at the structural transition, pointing towards a first-order nature of both phase transitions as stated in Refs. [24,25]. As outlined above, these ordering temperatures are close to, but not exactly at the values reported in literature [7,8,9,24,25]. The first order nature of both phase transitions is corroborated by the evaluation of the plateaus of the large pulse experiments yielding latent heats of about 70 J/mol and 220 J/mol for the magnetic and the Jahn-Teller transitions, respectively.

To get an estimate of the phonon contribution to the specific heat we calculated the phonon-derived heat capacity assuming one Debye term for the acoustic modes and formally assuming Einstein modes at three different eigenfrequencies to cover the remaining 36 degrees of freedom. A good fit of the heat-capacity results for temperatures above 50 K up to room temperature has been obtained using a Debye term of 205.5 K and Einstein frequencies corresponding to 195, 460, and 660 K (solid line in Fig. 5). A fit utilizing a similar set of parameters was also successful in describing the temperature-dependent heat capacity of GaV$_4$S$_8$ [23].

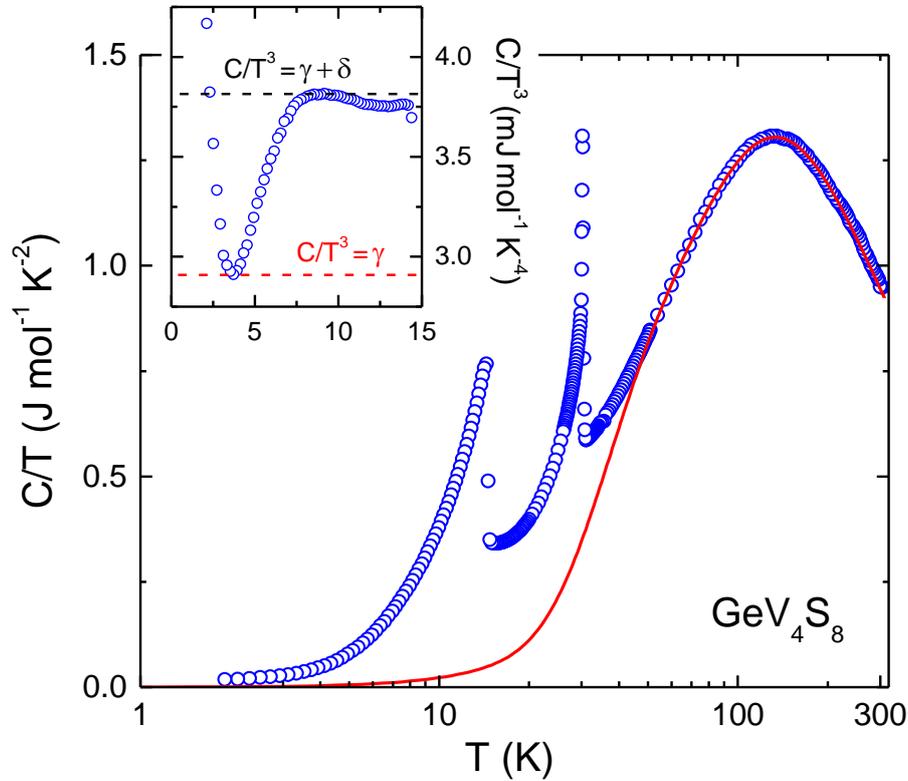

FIG. 5. Specific heat of GeV$_4$S$_8$ plotted as $C/T$ vs. a logarithmic temperature scale. The data have been collected on cooling. The solid line indicates the calculated phonon contribution. Inset: Low-temperature heat capacity plotted as $C/T^3$. The dotted lines indicate the two contributions to the low-temperature specific heat, where $\gamma$ corresponds to the pure phonon contributions, while the term $\gamma+\delta$ takes phonon and magnon contributions into account (see text for details). The $\gamma$ value of 2.91 mJ/mol K$^4$ results in a Debye temperature of 205.5 K.

To determine possible contributions of magnons to the heat capacity below the Néel temperature, the inset of Fig. 5 shows $C/T^3$ vs. temperature for temperatures between 2 K and $T_N$. In gapless antiferromagnets, the heat capacity at low temperatures is expected to follow $C = \gamma T^3 + \delta T^3$, where the first term describes phonon contributions and the second term is derived from gapless magnons with linear dispersion. The prefactor of the phonon term, $\gamma$, allows deducing the Debye temperature, $\theta_D$, while the prefactor $\delta$ is connected with details of the linear magnon dispersion and depends on the effective magnetic exchange or alternatively on the magnetic ordering temperature. Hence, in the representation of the inset of Fig. 5, we expect a constant dependence of the heat capacity being the sum of phonon and magnon terms. However, it seems that the heat capacity, plotted as $C/T^3$, towards 0 K extrapolates to the pure phonon-derived Debye temperature of $\theta_D$ = 205.5 K. This value was determined from the phonon fit as shown in Fig. 5. Hence, there are no magnon contributions at the lowest temperatures pointing towards a gapped antiferromagnet. The strong increase of the heat capacity $C/T^3$ towards low temperatures, $T <$ 4 K, probably either stems from nuclear contributions or, alternatively, from a linear glass-like term in the heat capacity. On increasing temperatures, for $T >$ 4 K magnon contributions evolve and exist up to the magnetic phase-transition temperature. This fact enables a rough estimate of the size of the antiferromagnetic magnon gap to be of the order of 0.5 meV. However, the increase around 5 K is much too steep to be explained by a purely exponential increase of the magnon-derived specific heat.

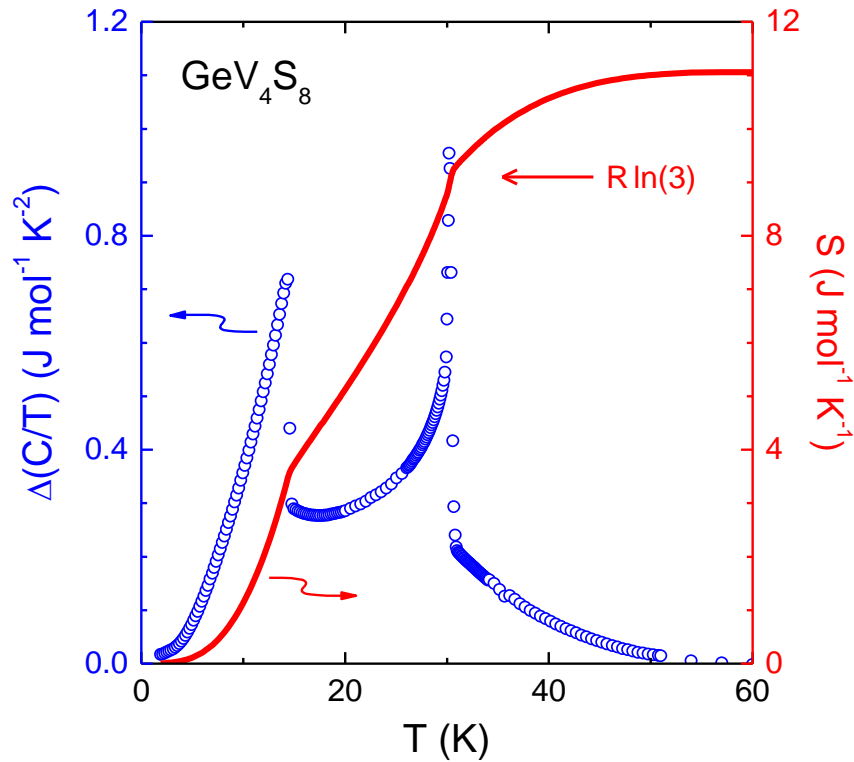

FIG. 6. Circles (left scale): Excess specific heat of GeV$_4$S$_8$ at the magnetic and structural phase transitions. The pure phonon contribution as indicated by the solid line in Fig. 5 has been subtracted from the experimental results. Line (right scale): Entropy release at the magnetic and structural phase transition. The arrow indicates the value of $R \ln(3)$, the entropy release that is expected at $T_N$ and is determined by the spin degrees of freedom alone.

The residual heat capacity $\Delta C$, plotted as $\Delta C/T$, which arises from the structural and magnetic phase transitions, is plotted in Fig. 6. Here the pure phonon-derived heat capacity, as estimated by the solid line in Fig. 5, has been subtracted from the measured data. In this representation, the area of this excess heat capacity provides a direct estimate of the entropy, released by the magnetic as well as by the structural phase transition (line in Fig. 6). For a spin $S = 1$ system, an entropy release of $R \ln(3) \approx 9.1$ J/mol K is expected at the magnetic transition. However, in GeV$_4$S$_8$ only less than 50% of this value show up at $T_N$ and the full entropy for the magnetic $S = 1$ system is reached just below the structural transition only. This fact documents that a large fraction of magnetic entropy is released in strong fluctuations well above the magnetic ordering temperature. This unusually strong release of spin entropy from the onset of magnetic order up to the structural phase transition indicates the significant coupling of spin, orbital, and lattice degrees of freedom between $T_N < T < T_{JT}$. The strong correlations between spin states and lattice distortions probably stem from the orbital degrees of freedom of the V$_4$ molecules, which start to order at $T_{JT}$, and strong magnetoelastic coupling effects have also been stressed in Ref. [25]. However, as discussed above, the magnetic and the structural transitions are of first order and the entropy release may strongly depend on the measurement mode.

In Fig. 7, we show the heat capacity at the structural [Fig. 7(a)] and the magnetic transition [Fig. 7(b)] on enlarged scales, measured at zero fields, compared to results at an external magnetic field of 9 T. To detect possible shifts of the transition temperatures by magnetic fields, these experiments have been performed with high temperature resolution. Astonishingly, it seems that both specific-heat anomalies are rather complex and seem to be composed of two subsequent transitions. In Fig. 7(a), at zero external magnetic field, the transition shows up as one well-defined lambda-like anomaly at 14.6 K. However, a significant two-step behavior of the heat capacity anomaly is found at 9 T, signaling two subsequent transitions in a narrow temperature window. While the onset of magnetic order is shifted to 14.5 K, a second transition appears close to 14.2 K. Maybe this is related to structural domains with different easy directions, where the onset of magnetic order depends on the orientation of the external magnetic field.

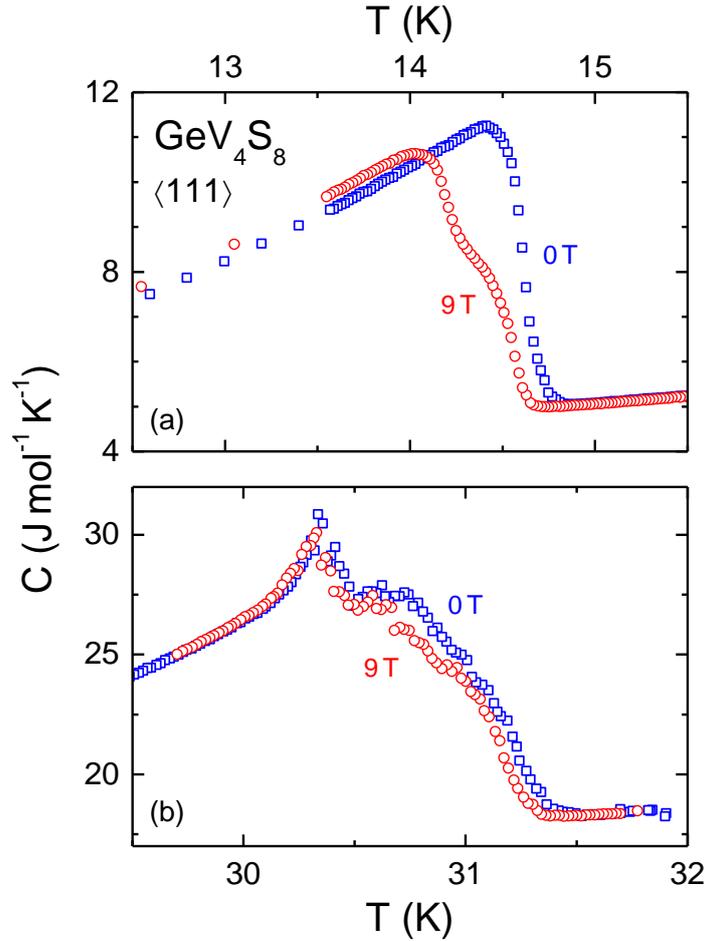

FIG. 7. Specific heat of GeV$_4$S$_8$ at temperatures close to the magnetic (a) and structural (b) phase transition. Measurements at zero external magnetic field are compared with measurements at 9 T. Both experiments were performed on cooling. The external magnetic field was oriented parallel to the crystallographic <111> direction.

As outlined above and reported in literature [25], GeV$_4$S$_8$ is characterized by strong magnetoelastic effects and the spin degrees of freedom are strongly coupled to the lattice. One could speculate that either the structural and magnetic transitions become decoupled at finite magnetic fields, or that the onset of collinear antiferromagnetic order with modulation along [½ ½ 0] at zero external field [8,24], in finite magnetic fields becomes replaced by two subsequent magnetic phase transitions, with more complex ordering wave vectors or spin patterns. The complexity of low-temperature magnetic phase diagrams in frustrated spinel lattices has recently been impressively documented in Ref. [32].

The situation is somewhat different in the high-resolution heat-capacity experiments around the Jahn-Teller transition, shown in Fig. 7(b). For both fields, these experiments reveal two, or even more subsequent transitions, which are located between 30.3 and 31 K and where the upper transition certainly is strongly smeared out. Both phase transitions show almost no shifts (< 0.1 K) in external magnetic fields of 9 T and appear to be extremely reproducible. At this Jahn-Teller transition, orbital order induces ferroelectricity and again, the onset of orbital and polar order might be decoupled. One possible explanation of this smeared out double-peak feature certainly is that it results from regions in the sample that are twinned, evidencing slightly different strain and, hence, revealing slightly different ordering temperatures. However, we will show later that these double structures indeed seem to be an intrinsic property of $GeV_4S_8$. The fact that these double transitions have not been detected in the temperature dependence of susceptibility and resistivity, probably stem from the fact that these techniques are not sensitive enough and that the temperature resolution was much lower compared to the specific-heat scans shown in Fig. 7.

### E. Thermal expansion

Figure 8 shows the thermal expansion of $GeV_4S_8$ as measured from the lowest temperatures up to 200 K. The measurements were performed along the crystallographic <111> direction in zero field and with an external magnetic field also applied along <111>. On cooling, the samples length $\Delta L/L$ continuously decreases, with a relative length change of 2 ‰, from $4\times10^3$ at room temperature (not shown) to $2\times10^3$ just above the Jahn-Teller transition. At the onset of orbital order, we observe a large step-like decrease of the sample dimension along the cube diagonal. The decrease appears abruptly, amounts to a change of macroscopic sample length of almost 4 ‰, and strongly reminds of a first-order phase transition. We should have in mind that in this orbital-ordering transition the $V_4$ tetrahedra are distorted, with one elongated V-V bond and one shortened bond at opposite sites. This change of bond lengths appears within the cube planes of the high-temperature fcc lattice and it is unclear how this distortion affects the <111> direction of the sample. Astonishingly, this length change of orbital ordering is partly compensated by an increase in length along <111> of approximately 2 ‰ at the onset of antiferromagnetic order. Again, this increase appears abruptly and step like, indicative of a first-order transition. From these thermal expansion experiments it is clear that in $GeV_4S_8$ spin and lattice degrees of freedom are strongly coupled. In accordance with the heat-capacity results, both transitions reveal only minor shifts in an external magnetic field of 9 T, with both transitions becoming slightly smeared out. As revealed by the inset of Fig. 8, for both phase transitions, there are no indications of two successive anomalies as found from the specific-heat measurements shown in Fig. 7. Obviously, the temperature resolution of the thermal expansion experiments is not high enough, or thermal-expansion is sensitive for structural aspects of the transitions, while the specific heat also senses the probably slightly decoupled spin ordering.

The thermal expansion experiments in $GeV_4S_8$ are only partly compatible with published results of diffraction experiments: The volume of the cubic phase at room temperature amounts 0.900 nm$^3$ [7], while the volume in the orbitally ordered phase is reduced to 0.898 nm$^3$ at 25 K [25]. This volume change amounts to approximately 2 ‰ and is only a small fraction of the effect which is expected on the basis of the thermal expansion. Assuming isotropic expansion effects only, the volume change should be of order 3 $\Delta L/L$ and compared to our results should be of order 1.8 %. Obviously, in $GeV_4S_8$ an isotropic volume reduction is a highly oversimplified assumption and much more detailed experiments are necessary to unravel these discrepancies. Based on diffraction experiments, it also has been concluded that antiferromagnetic order is established without further symmetry reduction but with a further enhancement of the existing distortion. Again, this does not seem not to be compatible with the results documented in Fig. 8, where the lattice distortion along

<111>, which appears at orbital ordering, becomes partly compensated with the onset of antiferromagnetic order.

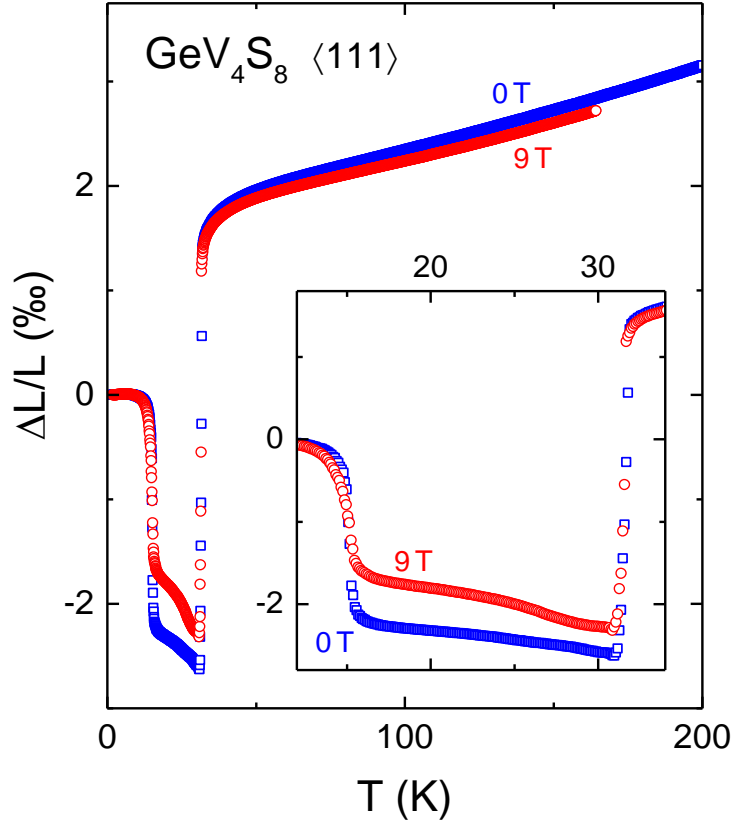

FIG. 8. Thermal expansion of GeV$_4$S$_8$ as measured along the crystallographic <111> direction in zero field and in a magnetic field of 9 T applied along <111>. The inset shows an enlarged region around the two phase transitions. These measurements have been performed on heating.

### F. Dielectric properties

Figure 9 shows the dielectric constant $\varepsilon'(T)$ of GeV$_4$S$_8$ measured at frequencies between 5.6 and 100 kHz along <111> and at temperatures around the two phase transitions. The magnetic and the structural phase transition are clearly visible in this plot and the dielectric constant reveals a rather unusual dome-like enhancement between the two phase transitions. The temperature dependence plotted in Fig. 9 was measured on heating. In the temperature range between the two phase transitions, enormous hysteresis effects appeared (not shown): On cooling, the temperature dependence of the dielectric constant is of similar shape, but exhibits much lower values of $\varepsilon'(T)$ in between the structural and magnetic phase transitions. Singh *et al*. [9] have reported very similar behavior. The highly unusual temperature and frequency dependence of the dielectric constant can be explained by the fact that at the structural phase transition strong dispersion effects arise, which dominate the dielectric response at radio frequencies, but which are fully suppressed again at the magnetic phase transition. In agreement with Ref. [9], we assume that ferroelectric order is induced by the appearance of orbital order at the structural phase transition. In the polar ordered phase, relaxational dynamics in the radio-frequency range dominates the dielectric properties. The second transition suppresses this dynamics again. It could be a transition into a paraelectric phase but also could be a transition into a second ferroelectric phase, with the polarization pointing along another symmetry direction of the crystal. From these experiments, it is clear that at orbital ordering, as well as at the magnetic phase transition, polar distortions of the lattice are strongly involved, in agreement with the findings of our thermal-expansion measurements (section IIE).

That the behavior at the structural phase transition is even more complex is documented in the inset of Fig. 9. Here the dielectric constant measured at 100 kHz is shown on an enlarged scale around the onset of the coupled orbital and polar order. On approaching the structural phase transition, the dielectric constant slightly increases, a fingerprint that the system comes close to polar order. A cusp-like anomaly and a strong decrease signal a structural phase transition close to 31 K. However, only about 1 K below the cusp temperature, $\varepsilon'(T)$ exhibits a minimum and strong dispersion effects start to appear as revealed in the main frame of Fig. 9. Again, similar to what was observed in the heat-capacity experiments (Fig. 7), the coupled orbital-order and polar phase transitions seem to be decoupled into two subsequent phase transitions. Experiments that are more systematic are necessary to unravel the nature of this highly unusual coupling of spin, orbital, and lattice degrees of freedom.

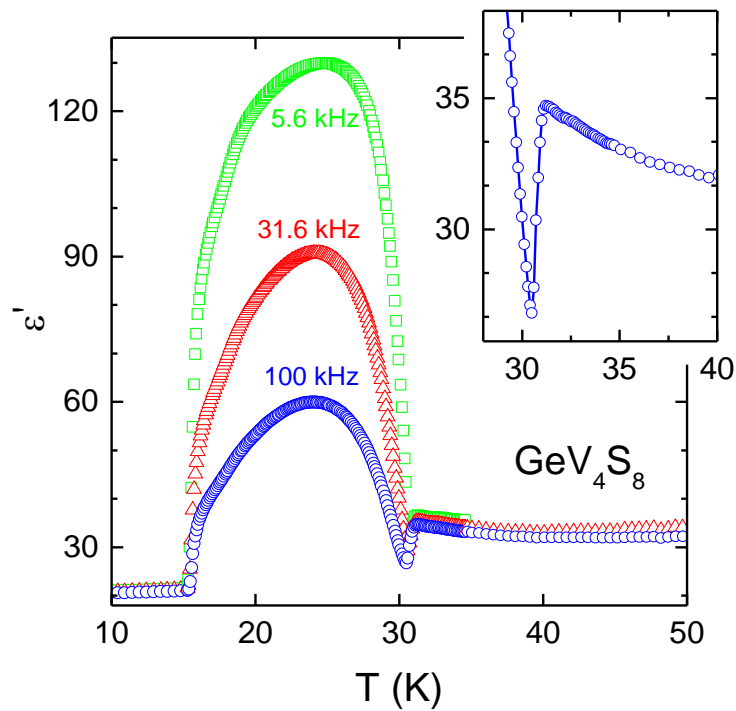

FIG. 9. Temperature dependence of the dielectric constant of $GeV_4S_8$ measured along the crystallographic <111> direction at frequencies between 5.6 and 100 kHz under heating. The inset shows an enlarged region around the onset of the coupled orbital and polar order.

## IV. SUMMARY AND CONCLUSIONS

In this manuscript, we have presented a detailed investigation of structural, electric, magnetic, thermodynamic, and dielectric properties of the lacunar spinel $GeV_4S_8$. The interest in this material stems from the fact that the lattice is built up of molecular units, with vanadium $V_4$ clusters characterized by spin $S$ = 1. Like all lacunar spinels, $GeV_4S_8$ undergoes an orbital-ordering transition at 30.5 K, which induces long-range ferroelectric order. However, in contrast to most lacunar spinels, it exhibits antiferromagnetic order below 14.6 K. The electrical resistivity increases by more than 14 decades down to the lowest temperatures. Close to room temperature, a gap energy of 0.23 eV is determined, close to the value expected from LDA + U calculations [8]. Below the Jahn-Teller transition, the temperature dependence of the resistivity can be described in terms of hopping conductivity. However, in this temperature range the resistivity reveals an unusually strong voltage dependence. The magnetic susceptibility can be described by a Curie-Weiss law with antiferromagnetic exchange and by a paramagnetic moment expected for a spin $S = 1$ system. The exchange interactions are of the order of -50 K and only slightly influenced by the onset of orbital order. In the heat-capacity experiments, two significant lambda-like anomalies show up at the structural, as well as at the magnetic transition. However, both transitions seem to be of first order. A closer inspection shows that both phase transitions could be more complex, revealing multiple transitions in a narrow temperature range. It is unclear if this is an intrinsic property of the sample or can be explained by a heterogeneously strained sample. The entropy at both transitions is significantly too low and between the magnetic and the structural transition, $GeV_4S_8$ is strongly dominated by magnetoelastic effects. The strong coupling of structural, orbital, and spin degrees of freedom also is documented by thermal expansion experiments. Finally, we found a highly unusual behavior in the temperature dependence of the dielectric constants, with strong dispersion effects only showing up between the magnetic and the ferroelectric transition. Again, we find some experimental evidence for a decoupling of orbital and polar phase transition.

      In summary, the lacunar spinel $GeV_4S_8$ is a cluster compound, characterized by tetrahedral vanadium units with total spin $S = 1$. At low temperatures, polar and magnetic order coexist, putting $GeV_4S_8$ into the large class of multiferroics. Orbital order at the Jahn-Teller transition induces ferroelectricity. Subsequently antiferromagnetic order is established at low temperatures. Specifically between the two phase transitions, spin, orbital, and lattice degrees of freedom are strongly coupled. Further investigations of phononic and magnetic excitations are highly warranted.

### ACKNOWLEDGMENTS

This research was supported by the DFG via the collaborative research center TRR 80 "From Electronic Correlations to Functionality" (Augsburg, Munich, and Stuttgart).